\documentclass[a4paper,twocolumn,11pt,unpublished]{quantumarticle}
\pdfoutput=1
\usepackage[utf8]{inputenc}
\usepackage[english]{babel}
\usepackage[T1]{fontenc}
\usepackage{amsmath}
\usepackage{hyperref}

\usepackage[numbers,sort&compress]{natbib}

\newcommand{\ket}[1]{\left|#1\right\rangle}
\newcommand{\bra}[1]{\left\langle #1\right|}

\usepackage{tikz}

\begin{document}

{\scriptsize This is the pre-peer reviewed version of the following article: A. A. Melnikov, L. E. Fedichkin, R.-K. Lee, and A. Alodjants, Adv. Quantum Technol. 3, 1900115 (2020), which has been published in final form at \href{https://doi.org/10.1002/qute.201900115}{doi.org/10.1002/qute.201900115}. This article may be used for non-commercial purposes in accordance with Wiley Terms and Conditions for Use of Self-Archived Versions.}\\\medskip

\title{Machine learning transfer efficiencies for noisy quantum walks}

\author{Alexey A. Melnikov}
\thanks{Corresponding author, e-mail: alexey.melnikov@unibas.ch, homepage: melnikov.info}
\affiliation{ITMO University, Kronverksky prospekt 49, 197101 St. Petersburg, Russia}
\affiliation{Valiev Institute of Physics and Technology, Russian Academy of Sciences, Nakhimovskii prospekt 36/1, 117218 Moscow, Russia}
\affiliation{Department of Physics, University of Basel, Klingelbergstrasse 82, 4056 Basel, Switzerland}

\author{Leonid E. Fedichkin}
\affiliation{Valiev Institute of Physics and Technology, Russian Academy of Sciences, Nakhimovskii prospekt 36/1, 117218 Moscow, Russia}
\affiliation{Moscow Institute of Physics and Technology, Institutskii pereulok 9, 141700 Dolgoprudny, Moscow Region, Russia}

\author{Ray-Kuang Lee}
\affiliation{Physics Division, National Center for Theoretical Sciences, 30013 Hsinchu, Taiwan}
\affiliation{Institute of Photonics Technologies, National Tsing Hua University, 30013 Hsinchu, Taiwan}

\author{Alexander Alodjants}
\affiliation{ITMO University, Kronverksky prospekt 49, 197101 St. Petersburg, Russia}

\maketitle

\begin{abstract}
Quantum effects are known to provide an advantage in particle transfer across networks. In order to achieve this advantage, requirements on both a graph type and a quantum system coherence must be found. Here we show that the process of finding these requirements can be automated by learning from simulated examples. The automation is done by using a convolutional neural network of a particular type that learns to understand with which network and under which coherence requirements quantum advantage is possible. Our machine learning approach is applied to study noisy quantum walks on cycle graphs of different sizes. We found that it is possible to predict the existence of quantum advantage for the entire decoherence parameter range, even for graphs outside of the training set. Our results are of importance for demonstration of advantage in quantum experiments and pave the way towards automating scientific research and discoveries.
\end{abstract}

\section{Introduction}

Classical and quantum particle transport plays a significant role in many scientific fields related to the transfer of charge~\cite{giese2001direct}, energy~\cite{fleming2007photosynthesis,fleming2009photosynthesis,engel2010photosynthesis} or information~\cite{farhi1998qw}. Known advantages of coherently propagating quantum excitations are used for developments in quantum computing~\cite{Childs1,Childs2}, search algorithms~\cite{childs2004search,yasser2016search,leonardo2018search}, communication networks, and efficient energy transport~\cite{engel2007evidence,mohseni2008environment}. Understanding the origins of quantum transport advantage is important for these fields.
The standard approach for studying quantum transport phenomena is based on the quantum walks model~\cite{aharonov1993qw}. Using this model it was shown that quantum particles propagate faster than classical on certain graphs including line~\cite{ambainis2001one}, cycle~\cite{solenov2006continuous,fedichkin2006mixing}, hypercube~\cite{kempe2005discrete,PhysRevA.73.032341}, and glued trees~\cite{Childs:2003:EAS:780542.780552} graphs. A systematic study of arbitrary graphs has several challenges. The number of possible graphs to study grows as the factorial of the number of graph vertices~\cite{slone1964online}, although new ways to reduce the problem dimensionality are being developed~\cite{novo2015systematic}. 

To overcome this challenge, we developed an automated approach to study quantum transport properties and to predict the possibility of quantum advantage in particle transfer. This automated approach is based on using a specific binary classifier called classical-quantum convolutional neural network (CQCNN), recently introduced in ref.~\cite{melnikov2019predicting}, which learns to predict whether the quantum or classical transport is more efficient on a graph with a given topology. 
However, it is known that decoherence is inevitable in quantum systems. Under which levels of noise can we still expect a quantum advantage to hold? To answer the question, one needs to perform simulations for all graphs of interest, testing different levels of noise. The noise is changing the quantum dynamics in a non-monotonic way, sometimes helping quantum particles to reach the target node faster~\cite{kendon2003decoherence,plenio2016decoherence}. To find out the exact relations, one would need to simulate the dynamics for all levels of decoherence.

In this paper, we demonstrate how the study of noisy quantum walks can be automatically performed by a neural network that learns from restricted simulated dynamics. Our approach is based on using a version of CQCNN that is augmented with a capacity to learn from additional data about decoherence levels. The new approach is tested on a family of cycle graphs, which represent a specific interest in quantum transport studies. We simulated a set of cycle graph examples and observed that the developed neural network can find correct conditions on the decoherence levels for graphs never given to the neural network.

\section{Classifying noisy quantum walks}

Quantum transport and a corresponding classical transport are modeled by stochastic processes of quantum walks~\cite{aharonov1993qw,kempe2003qw,venegas2012quantum}, and classical random walks~\cite{lovsz2002random,masuda2017random}, respectively. More specifically, the state of a quantum particle in a graph defined by adjacency matrix $A$ (or transition matrix $T$) is described by a density matrix $\rho$, which evolves according to the Gorini–Kossakowski–Sudarshan–Lindblad equation~\cite{GKS1976, L1976,manzano2019short}
\begin{align}\label{qwdynamics}
  &\frac{\mathrm{d}\rho(t)}{\mathrm{d}t} = - \frac{i}{\hbar}(1-p)\left[\mathcal{H},\rho(t)\right]\nonumber\\
   &+ p\sum_{mk}\left( L_{mk}\rho(t) L_{mk}^\dagger - \frac{1}{2}\left\{L_{mk}^\dagger L_{mk},\rho(t)\right\}\right)\nonumber\\
   &+ \gamma\left(L_s\rho(t)L_s^\dagger -\frac{1}{2}\left\{ L_s^\dagger L_s, \rho(t)\right\}\right). 
\end{align}
The Hamiltonian $\mathcal{H}=\hbar A$ defines coherent continuous time transitions of a particle on the graph, $L_{mk}=T_{mk}\ket{m}\bra{k}$ and $L_{s}=\ket{s}\bra{t}$ operators correspond to transitions from vertices $k$ to $m$ and from $t$ (target) to $s$ (``sink''), respectively. The ``sink'' vertex is an additional vertex, which is coupled to the target vertex, and is constantly monitored for the presence of a particle. The sink vertex is an important addition because a continuous measurement directly in the target vertex might lead to the undesired quantum Zeno effect.

The parameters $\gamma$ and $p$ further define the quantum walk dynamics: $\gamma$ is the coupling of the target vertex to the sink vertex, and $p$ is the decoherence parameter. In particular, depending on the value of $p$ the transport can be either quantum ($p=0$), or classical ($p=1$, no sink vertex) that is defined by the probability distribution
\begin{equation}\label{rwdynamics}
    \pi(t) = \mathrm{e}^{-t}\mathrm{e}^{Tt}\pi(0).
\end{equation}
In the classical case the sink vertex is not needed, because the measurement procedure does not affect the state of the particle. The coupling to the sink vertex is set to the value of $\gamma=1$ throughout this paper.

From a physical interpretation of Eq.~(\ref{qwdynamics}), particle transitions between vertices can be recognized as tunneling processes with temperature-dependent coefficients. In this case, Eq.~(\ref{qwdynamics}) is inherent to the dissipative tunneling problem for a physical system established by a graph of some topology. It is important that some fundamental physical properties of this problem are well-known analytically, but in the two-site limit only~\cite{leggett1983quantum,larkin1983quantum,grabert1984quantum,grabert1987quantum}. In particular, there exist some temperature of crossover, or, a phase transition from classical to quantum regimes. In ref.~\cite{Alodjants2017} one of us has shown that this kind of phase transition reduces to hybridization of quantum algorithms based on quantum tunneling processes.  
The microscopic description of the temperature-dependent tunneling process in the presence of dissipation requires characterization of interaction with a reservoir that needs a separate analysis, cf. refs.~\cite{leggett1983quantum}. In this paper, we restrict ourselves by the simplified model of decoherence established by Eq.~(\ref{qwdynamics}).

Solutions to Eqs.~(\ref{qwdynamics}) and~(\ref{rwdynamics}) directly provide probability distributions of particle's position in the graph defined by $A$, and given the value of $p$. In the particle transport problem we are interested in the probability of finding a particle in the target (or, in the quantum case, sink) vertex. If this probability is larger than $1/\log n$, where $n$ is the total number of vertices, it is assumed that the particle has reached the target. Hence, by comparing the solutions to Eqs.~(\ref{qwdynamics}) and~(\ref{rwdynamics}), we can define the particle transfer efficiency: it is $1$ if the quantum particle reached the target first, and $0$ otherwise.

\begin{figure}[t!] 
  \centering
  \includegraphics[width=1\linewidth]{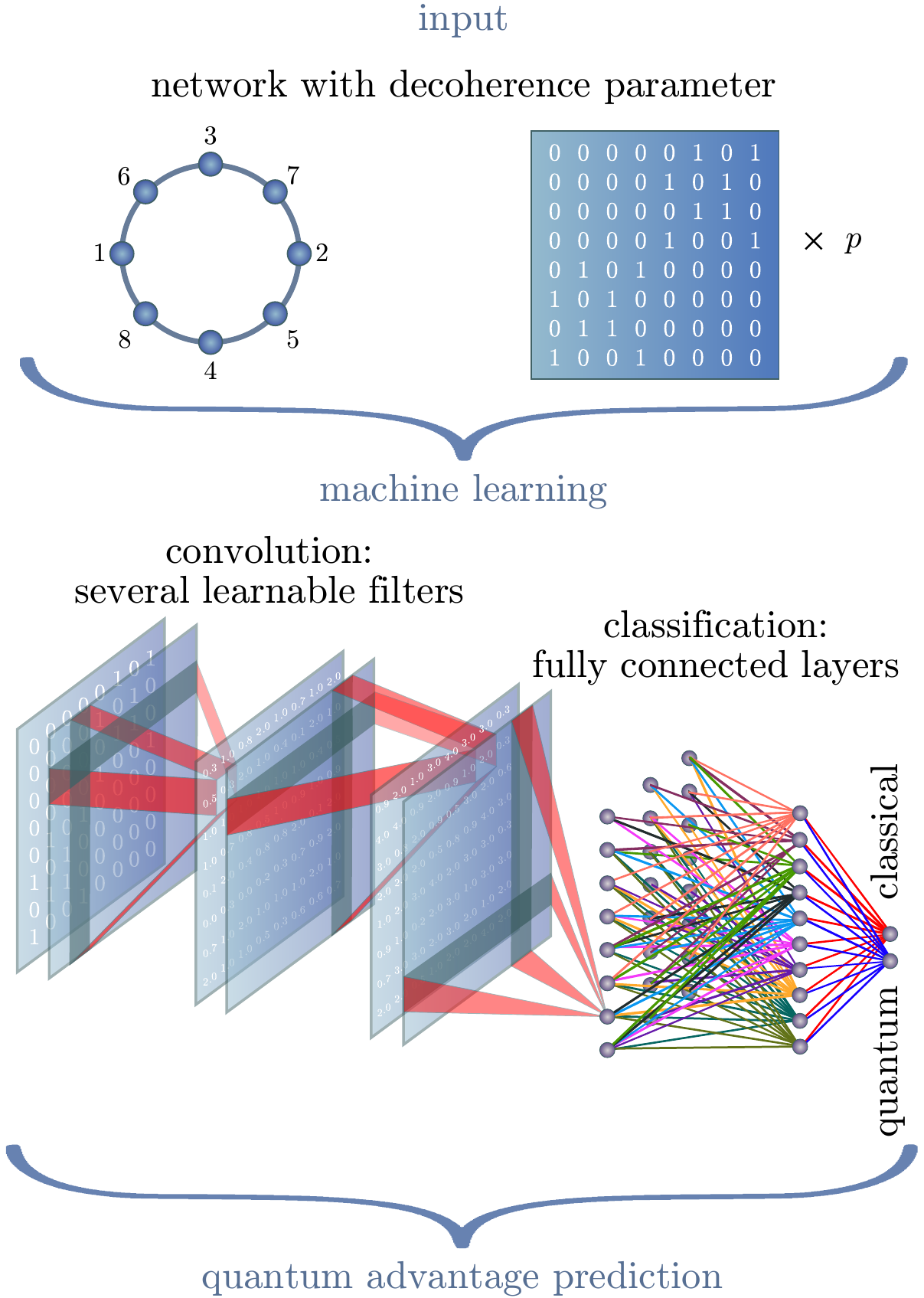}
  \caption{A scheme of the machine learning approach that is used to determine the exact conditions under which the quantum transfer efficiency advantage can be expected. The approach is based on using the convolutional neural network. One out of three levels of filters is shown.}
  \label{fig:figure1}
\end{figure}

Determining the transfer efficiency usually requires the full numerical simulation of Eq.~(\ref{qwdynamics}) for every given adjacency matrix $A$ and parameter $p$, as described above. In this work, we show that this is not always required. To predict the result of the dynamics from Eq.~(\ref{qwdynamics}) we use the supervised learning approach shown in Figure~\ref{fig:figure1} that schematically demonstrates the working principle of quantum advantage prediction. In this machine learning approach, a network on which the dynamics was simulated is given as a training example. The form of the input to the neural network is an adjacency matrix $A$ and the decoherence parameter $p$. This input is processed by a convolutional neural network~\cite{krizhevsky2012imagenet} called CQCNN~\cite{melnikov2019predicting} with specifically designed learnable ``cross'' filters. The first layer of ``cross'' filters extracts features corresponding to a function of weighted numbers of neighboring edges (for a detailed mathematical description see ref.~\cite{melnikov2019predicting}). The second layer of ``cross'' filters extracts information about neighboring edges of neighboring edges. The third layer continues to learn deeper about neighbors of neighbors, and passes information about vertices connectivity further. The described filters are shown in Figure~\ref{fig:figure1} as semitransparent squares with highlighted columns and rows: they take values from the previously processed layer (squares with values) which starts with the matrix $A$ at the very first level. All these convolutions are followed by fully connected layers of neurons, which are shown in Figure~\ref{fig:figure1} as small balls fully connected to each other between the layers. There are three layers of neurons with $3n$, $10$, and $2$ neurons in each layer, respectively. After processing the graph in these layers, the neural network gives its prediction on quantum advantage in efficient transport. The training data is used to compute the loss using the cross-entropy loss function, and optimize CQCNN's weights by the stochastic gradient descent optimization technique. The trained network, as we show next, can predict quantum advantage on a graph without being trained with data from this graph.

\section{Classifying unknown transport \\dynamics on cycle graphs}
\begin{figure}[t!] 
  \centering
  \includegraphics[width=0.85\linewidth]{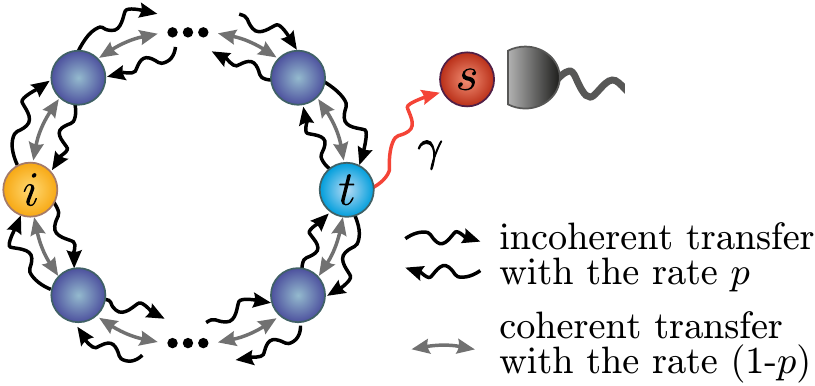}
  \caption{A schematic representation of a cycle graph with an arbitrary number of vertices. The vertices are connected in different ways defined by Eq.~(\ref{qwdynamics}).}
  \label{fig:figure2}
\end{figure}

As an example of classifying and predicting unknown noisy quantum walk dynamics we consider cycle graphs. Cycle graphs are known to provide a speedup for one- and two-particle quantum walk in mixing and hitting time~\cite{solenov2006continuous,fedichkin2006mixing,melnikov2019hitting}. This advantage can moreover be used for quantum information purposes~\cite{melnikov2016quantum}. A cycle graph is schematically shown in Figure~\ref{fig:figure2} with an additional detector that is used to measure particle's state in the sink vertex $s$. Each vertex represents a possible position of a particle and is shown as a colored circle. The particle is initially placed in the vertex $i$ (yellow), and then moves according to Eq.~(\ref{qwdynamics}). All terms of Eq.~(\ref{qwdynamics}) are schematically visualized in Figure~\ref{fig:figure2}. The first term in Eq.~(\ref{qwdynamics}) is represented by gray arrows, the second term is represented by black arrows, whereas the third term is the red arrow. The goal is to reach the target vertex $t$ (blue), which is connected to a sink vertex $s$ (red) located near the particle detector. One can see that all three processes occur with different frequencies: $(1-p)$ for the coherent particle transfer, $p$ for the incoherent transfer, and $\gamma$ for the measurement procedure.

\begin{figure}[t!] 
  \centering
  \includegraphics[width=1\linewidth]{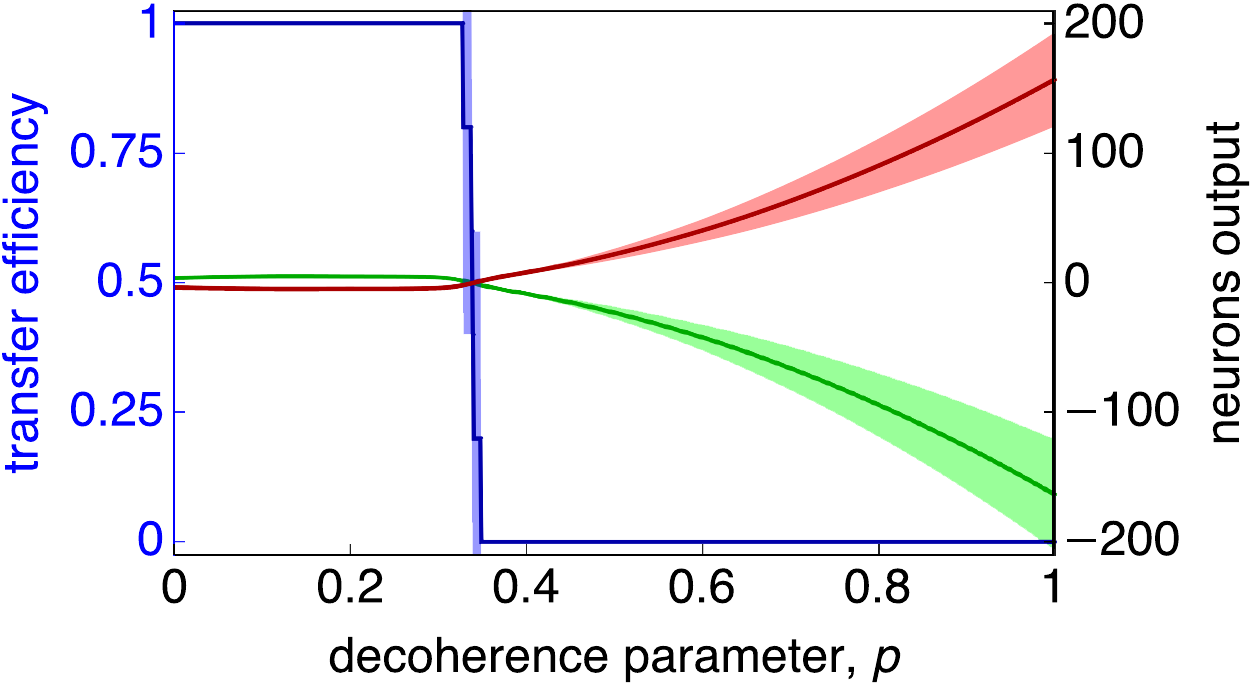}
  \caption{Prediction of transfer efficiency (blue) for a $6$-cycle graph for different values of decoherence parameter $p$. The activation values of output neurons are shown in red and green. The results are an average of $5$ CQCNN networks. Shaded areas correspond to the standard deviation.}
  \label{fig:6cycle}
\end{figure}

To study the effect of decoherence on particle transfer efficiency, we first simulate quantum walk dynamics on a $6$-cycle and measure the efficiency of transport between opposite vertices of the graph. The quantum walk and random walk dynamics are simulated for $1000$ randomly sampled values of the decoherence parameter $p$. Then, the results of these simulations are used to train CQCNN. Once the neural network is trained, we ask the network to predict if the quantum walk can lead to an advantage for a new given parameter $p$. The results of the transfer efficiency predictions are shown in Figure~\ref{fig:6cycle} as a blue line. The efficiency of one corresponds to quantum advantage in transport, whereas the efficiency of zero corresponds to a classical transport regime. We see that for $p<0.34$, the quantum transport is more efficient (efficiency of $1$), whereas for $p>0.34$, the quantum transport is less efficient (efficiency of $0$) than classical transport.

The crossover from quantum to classical transport occurs at $p=0.34$ and could be inherent to a second-order phase transition from quantum to classical tunneling that happens for a complex graph system at some finite temperature (note that the parameter $p$ is, in general, temperature-dependent)~\cite{larkin1983quantum,grabert1984quantum,grabert1987quantum}. Importantly, the efficiency is defined relative to the coupling parameter $\gamma$ and in case of properly chosen $\gamma$ the quantum transport can be at least as efficient as the classical transport.
Predictions of CQCNN are based on the learned values of the output neurons, which are shown in red (``classical'' class) and green (``quantum'' class) in Figure~\ref{fig:6cycle}. The decision about the class is made by the maximum value of the output neurons activation. We can see that the ``vote'' for the quantum class grows up to a particular value with the maximum at about $p=0.2$, which corresponds to the highest confidence for the quantum class. After the crossover point of $p=0.34$, the confidence in the classical class grows rapidly, which means the separation between classes become more apparent the more decoherence grows.

\begin{figure*}[ht!]
  \centering
  \includegraphics[width=0.9\linewidth]{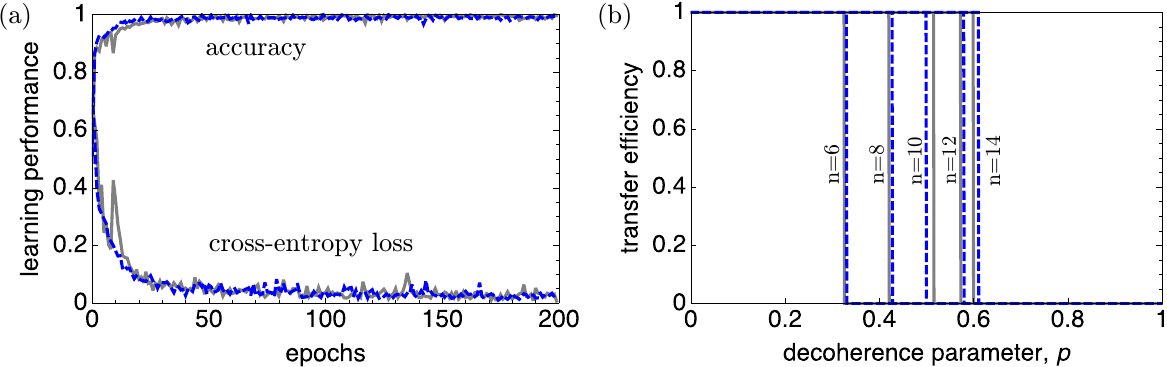}
  \caption{(a) Learning performance of the machine learning approach. Loss on the training data goes down to zero with the number of epochs. The accuracy of neural network predictions improves with the number of epochs and goes up to unit accuracy. (b) Prediction of transfer efficiency for cycles graphs of different sizes with $n=6,8,10,12$ and $14$ vertices. Dashed blue lines show the efficiency predicted by CQCNN after being trained on all graphs, whereas solid gray lines are obtained from CQCNN not trained on $10$-cycle graphs.}
  \label{fig:figure4}
\end{figure*}

We next use the neural network CQCNN of the same architecture as shown in Figure~\ref{fig:figure1}, and train it with simulations of noisy quantum walks on different graphs with $100$ data points each. After training, all graphs are tested on a combined set of $6,8,10,12$ and $14$-cycle graphs.
Figure~\ref{fig:figure4}(a) demonstrates the learning performance of two types of CQCNNs: one is trained on all graphs (gray), and the other is trained on all but $10$-cycle graphs (blue). One can see that both types of CQCNNs show very similar performance, suggesting that CQCNN has a generalization ability and does not need to be trained on all graphs. 
To verify that the predictions are accurate for all studied cycle graph dimensions, we check transfer efficiencies as functions of the decoherence parameter for each graph individually. The dependencies are shown in Figure~\ref{fig:figure4}(b), where one can see $5$ transitions from quantum-enhanced transport to classical transport that correspond to $5$ different cycle graphs. One can see that the match between gray and blue lines is very precise for $6,8$ and $12$-cycle graphs and less accurate for $10$ and $14$-cycle graphs, although we believe the match will increase with the number of training examples.

\section{Conclusion and Outlook}

The transfer of quantum and classical particles from a classification perspective was studied. The problem we considered is finding out which graphs, and under which conditions on decoherence, can provide a quantum advantage. This is especially relevant for near-future experimental demonstrations of quantum-enhanced transport, e.g., in lossy photonic, or polaritonic tunnel-coupled waveguides~\cite{Alodjants2017}. Physically, the crossover from quantum to classical transport may be relevant to a study of phase transitions from quantum to classical tunneling that occurs in complex graph systems at some finite temperature. In this paper, we developed a machine learning approach that can predict if a quantum advantage is possible for a given graph under a given noise level. The approach is based on training a convolutional neural network, called CQCNN, that automatically learns to extract feature vectors from graph adjacency matrices combined with a decoherence parameter. We demonstrated that not only CQCNN can find parameter range for which advantage holds on a given graph, but also for graphs which were not observed before. 
These results highlight the successful feature extraction from the simulated noisy quantum walk dynamics, which goes towards an understanding of the nature of quantum advantage. The presented machine learning approach helps in further developing quantum experiments~\cite{dunjko2018machine,melnikov2018active,wallnofer2019machine} showing an advantage of quantum transport.\\

\section*{Acknowledgements}
This work was financially supported by the Government of the Russian Federation, Grant 08-08, and by RFBR grants No. 19-52-52012 MHT-a and No. 17-07-00994-a. A.A.M. acknowledges funding by the Swiss National Science Foundation (SNSF), through the Grant PP00P2-179109. R.K.L. acknowledges funding by the Ministry of Science and Technology of Taiwan (No. 105-2628-M-007-003-MY4, 108-2627-E-008-001, 108-2923-M-007-001-MY3). The investigation was supported by Program No. 0066-2019-0005 of the Ministry of Science and Higher Education of Russia for Valiev Institute of Physics and Technology of RAS.

\section*{Conflict of Interest}
The authors declare no conflict of interest.


\end{document}